\documentclass{article}

\usepackage{arxiv}
\usepackage{multirow}
\usepackage{algpseudocode}
\usepackage[T1]{fontenc}    
\usepackage{hyperref}       
\usepackage{url}            
\usepackage{booktabs}       
\usepackage{amsfonts}       
\usepackage{nicefrac}       
\usepackage{microtype}      
\usepackage{lipsum}		
\usepackage{caption}
\usepackage{subcaption}
\usepackage{amsmath}
\usepackage{graphicx}
\usepackage[ruled,vlined]{algorithm2e}
\usepackage[square,numbers]{natbib}
\usepackage{doi}
\usepackage{mwe}
\usepackage{array}
\newcolumntype{P}[1]{>{\centering\arraybackslash}p{#1}}
\title{Reducing Network Cooling Cost using Twin-Field Quantum Key Distribution}

\author{ \href{https://orcid.org/0000-0000-0000-0000}{\includegraphics[scale=0.06]{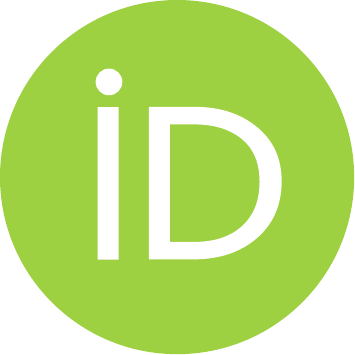}\hspace{1mm}Vasileios Karavias} \\
	Department of Physics\\
    University of Cambridge\\
	Cambridge, UK, CB3 0HE \\
	\texttt{vk330@cam.ac.uk} \\
	\And
    \href{https://orcid.org/0000-0000-0000-0000}{\includegraphics[scale=0.06]{orcid.pdf}\hspace{1mm}Andrew Lord} \\
    BT, Adastral Park\\ Martlesham Heath\\
    Ipswich, UK, IP5 3RE
	\And

	\href{https://orcid.org/0000-0000-0000-0000}{\includegraphics[scale=0.06]{orcid.pdf}\hspace{1mm}Mike Payne} \\
	Department of Physics\\
    University of Cambridge\\
	Cambridge, UK, CB3 0HE \\
}

\renewcommand{\shorttitle}{Reducing Network Cooling Cost using Twin-Field Quantum Key Distribution}

\hypersetup{
pdftitle={A template for the arxiv style},
pdfsubject={q-bio.NC, q-bio.QM},
pdfauthor={David S.~Hippocampus, Elias D.~Striatum},
pdfkeywords={First keyword, Second keyword, More},
}

\begin{document}
\maketitle
\begin{abstract}
Improving the rates and distances over which quantum secure keys are generated is a major challenge. New source and detector hardware can improve key rates significantly, however it can require expensive cooling. We show that Twin-Field Quantum Key Distribution (TF-QKD) has an advantageous topology allowing the localisation of cooled detectors. This setup for a quantum network allows a fully connected network solution, i.e. one where every connection has non-zero key rates, in a box with sides of length up to $110km$ with just $4$ cooled nodes, while Decoy state BB84 is only capable of up to $80km$ with $40$ cooled nodes, and $50km$ if no nodes are cooled. The average key rate in the network of the localised, cooled TF-QKD is $>30$ times greater than the uncooled Decoy BB84 solution and $\sim 0.9$ those of cooled Decoy BB84. To reduce the cost of the network further, switches can be used in the network. These switches have losses ranging between $1-2dB$. Adding these losses to the model shows further the advantages of TF-QKD in a network. Decoy BB84 is only able to generate fully connected solutions up to $20km$ if all nodes are cooled for a $40$ node network for $1dB$ losses. In comparison, using TF-QKD, $70km$ networks are possible with just 4 cooling locations for the same losses. The simulation shows the significant benefits in using TF-QKD in a switched network, and suggests that further work in this direction is necessary.

\end{abstract}


\section{Introduction}
Quantum Key Distribution (QKD) has been the subject of intense study in recent years due to the ability to generate information theoretic secure keys. In the absence of quantum repeaters, many protocols have been proposed to maximise the secure key generation rates and distances, from the original BB84 \citep{Bennett2014} to SARG04 \citep{Scarani2004}, Decoy State BB84 \citep{Hwang2003, Lo2005} and many more \citep{Mu1996, Tamaki2014, Stucki2009}. Simultaneously, the hardware required to carry out QKD has also improved, with quantum dot technologies being developed for high quality single and entangled photon sources \citep{Senellart2017, Wang2019}, superconducting nanowire single photon detectors (SNSPDs) \citep{Kahl2015, Zhang2019} being developed and commercialised as improvements on the single photon avalanche diode (SPAD) detectors \citep{Vines2019} as well as low loss fibres \citep{Tamura2018}. The combination of these have allowed secure keys to be generated at record distances of 421km \citep{Boaron2018}.\par
However, the protocols described above suffer not only from limited distance but also are vulnerable to attacks on the sources and detectors. To counter this, device-independent QKD protocols have been devised \citep{Pironio2009}, however they have challenging hardware requirements and suffer from low key rates. Measurement-Device-Independent QKD (MDI-QKD) \citep{Lo2012, Pirandola2012} improved on the feasibility of these protocols, being secure against attacks on the detectors, where most types of attacks are possible \cite{Lo2014}. However, the protocol still suffers from relatively strict hardware requirements and low rates. Recently, Twin-Field QKD (TF-QKD) \citep{Lucamarini2018} was proposed as a measurement device-independent protocol with high rates and long distance key generation capabilities, that beats the PLOB bound \cite{Pirandola2017}, thus extending the working range to the single untrusted repeater regime. It achieved this by replacing the Bell measurement in MDI-QKD with single photon interference. There have been many extensions to the original TF-QKD, such as Phase Matched QKD \cite{Ma2018}, sending-not-sending TF-QKD \cite{Wang2018, Xu2020} and many others \cite{Cui2019, Lin2018, Curty2019} with the aim of further improving the rates of the protocol. There have been many experimental implementations \citep{Minder2019, Chen2021}, and recently secure keys have been generated at distances of over 600km \citep{Pittaluga2021}. Here, we focus on the protocol proposed by \cite{Curty2019} and its extension to asymmetric channels by \cite{Wang2020}.  \par

\begin{figure}[t]
   \centering
    \includegraphics[width=0.6\linewidth]{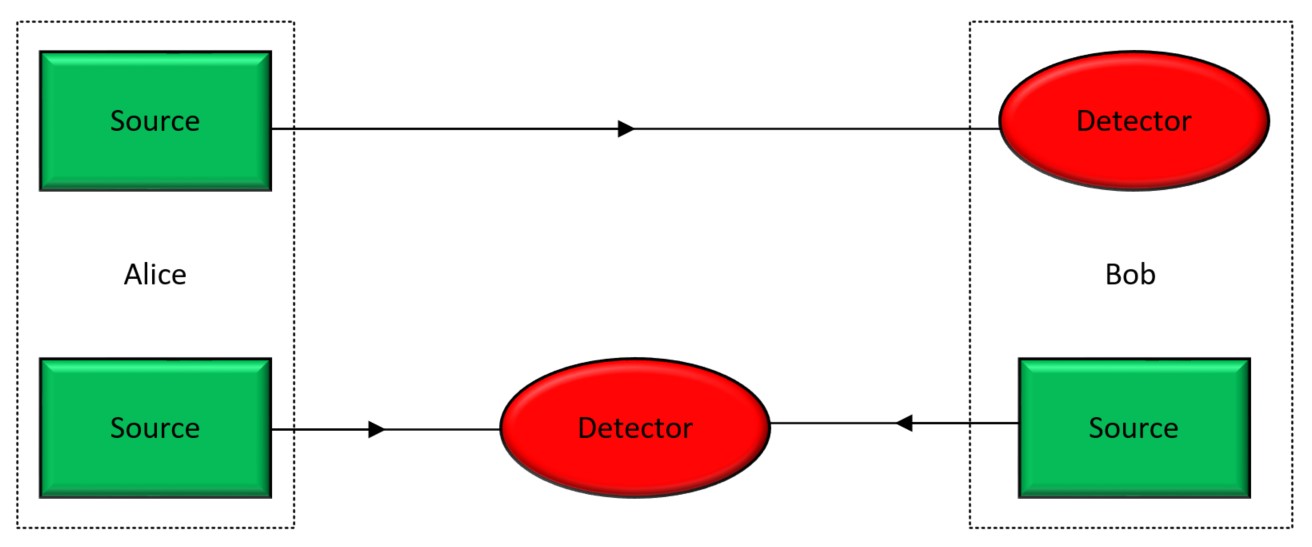}
    \caption{Topological Differences in setup for BB84, SARG04, Decoy State BB84 (top) compared to TF-QKD (bottom). In the top topology the source and detectors are fixed at the endpoints of communication whereas in the bottom topology the detector is not fixed allowing for easy collocation of detectors in the network}
    \label{fig:figure1}
\end{figure}
Figure \ref{fig:figure1} shows the difference in topologies of a system generating keys using the Decoy BB84 protocol compared to using TF-QKD. While the advantages of using TF-QKD in a point-to-point connection have been discussed, the advantage that TF-QKD brings in large quantum networks has not been studied. In this paper, we will demonstrate the advantage in a network of a topology where the detector is not fixed at the endpoint of the connection. In particular, we will show that TF-QKD with all detectors localised on just a few cooled locations can achieve similar rates and improved range as using Decoy BB84 where all the nodes in the network contain cooled SNSPDs in an optical switching paradigm. The localisation of cooling allows for an economically viable implementation benefiting from high quality SNSPDs allowing for secure key generation in networks larger than previously possible.

\begin{table}[t]
    \centering
    \begin{tabular}{|c|c|}
        \hline Parameter & Value \\
        \hline
        Mean photon number $\mu$ & 0.1 \\
        \hline
        Fibre loss $\alpha$ & 0.2~dB/km\\
        \hline
        Pulse repetition rate $f_{rep}$  & 100~MHz\\
        \hline
        \multirow{2}{*}{Dark count rate \citep{Quantique2021a, Quantique2021b} $r_{DC}$}  & 100~Hz (cold)\\
        & 200~Hz (hot)\\
        \hline
        \multirow{2}{*}{Detector efficiency \citep{Quantique2021a, Quantique2021b} $\eta_d$} & 0.85 (cold) \\
        & 0.2 (hot) \\
        \hline
        \multirow{2}{*}{Dead time \citep{Quantique2021a, Quantique2021b} $\tau_{dead}$}  & 1~$\mu$s (cold)\\
         & 50~$\mu$s (hot)\\ \hline
    \end{tabular}
    \caption{Photon source, fibre and photon detector properties.}
    \label{tab:variables}
\end{table}
\begin{figure}[t]
   \centering
    \includegraphics[width=0.6\linewidth]{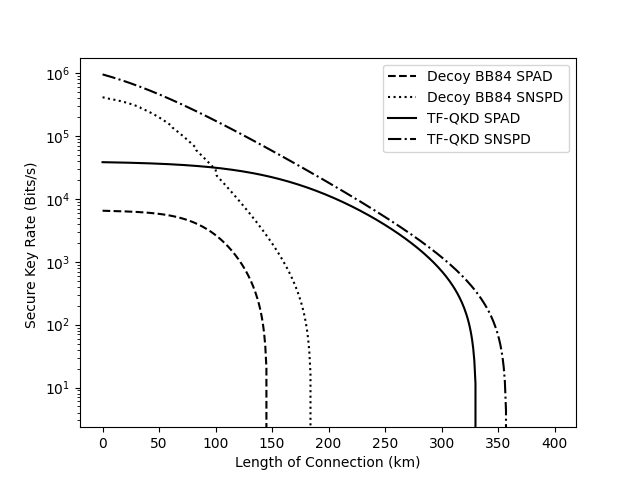}    \caption{Secure key rates calculated using equations (1) and (2) with parameters given in table \ref{tab:variables}. TF-QKD rates are shown for a symmetric setup.}
    \label{fig:figure2}
\end{figure}

\section{Secure Key Rates}
QKD secure key rates for the Decoy state BB84 and TF-QKD protocols are modelled using a Poissonian source, an optical fibre and photon detectors whose relevant parameters are tabulated in Tab.~\ref{tab:variables}. For a given length $l$ of fibre, define the fibre efficiency as $\eta_f = 10^{-\frac{\alpha l}{10}}$. A 3.5~ns detector window is used giving a dark count probability of $p_{DC}=r_{DC}\times 3.5~\text{ns}$.\par
The secure key rate for the Decoy state BB84 is \cite{Lo2005} 
\begin{equation}
    R = -Q_{\mu}f(E_{\mu})H_2(E_{\mu}) + Q_1[1-H_{2}(e_1)]
\end{equation} where $H_2(x) = - x \log x - (1-x) \log 1-x$ is the binary entropy, and the efficiency of the error correction code compared to the Shannon limit \cite{Eraerds2010} $f(E_{\mu}) = 1.2$ in our model. The overall gain $Q_{\mu}$ can be obtained using $$Q_{\mu} = p_{\mu}\eta_{dead} ~ : p_{\mu} = (\sum_{n=0} Y_n \frac{e^{-\mu}\mu^n}{n!})$$ where the n photon yield \cite{Fung2006, Lo2005}$$Y_n = [1-(1-\eta_f\eta_d)^n] + [1-\eta_f\eta_d]^n p_{DC}$$ The fraction of photons lost due to the detector dead time\cite{Eraerds2010} $\eta_{dead} = (1+\tau_{dead}f_{rep}p_{\mu})^{-1}$ and the single photon gain $Q_1$ can be obtained using \cite{Lo2005} $$Q_{1} = Y_1\mu e^{-\mu} \eta_{dead}$$The quantum bit error rate (QBER) in single photon states $e_1$ can be obtained by fitting $e_n$ to \cite{Lo2005} 
\begin{equation*}
Q_{\mu}E_{\mu} = (\sum_{n=0} Y_n \frac{e^{-\mu}\mu^n}{n!}e_n)\eta_{dead}
\end{equation*}
for the different values of $\mu$ used as decoys.\par
For BB84 the QBER is  $E_{\mu} = (1-V)/2$ for a visibility $V$ which is approximated by 
$$
    V = \frac{\mu \eta_f\eta_{d}}{\mu \eta_f \eta_{d} + 2P_{e}},
$$
with $P_e$  modelled as $P_e = (5.3 \times 10^{-7} + p_{DC})$\cite{Gobby2004}.\par

For the Twin-Field QKD, the secure key rate is given by \cite{Curty2019, Wang2020}
\begin{equation}
    R_X^{low} = \max \{R_{X,10}^{low}, 0\} + \max \{R_{X,01}^{low}, 0\}
\end{equation}

$$
   R_{X,k_ck_d}^{low} = p_{XX}(k_c,k_d)[1-H_2(e_{X,k_ck_d}) \\- H_2(\min\{\frac{1}{2}, e_{Z,k_ck_d}^{upp}\})] 
$$
where $p_{XX}(k_c, k_d)$ is the conditional probability on the announcement outcomes of Charlie given Alice and Bob both generate a signal state. $e_{X, k_ck_d}$ is the bit error rate of the system. $e_{Z,k_ck_d}^{upp}$ is the upper bound phase error rate of the system. We follow the methods in \cite{Wang2020} to calculate these values accounting for detector and source inefficiencies and misalignment with the same parameters defined above. See Appendix A for details of the calculation. The resulting secure rates are plotted in Figure \ref{fig:figure2} for both SPADs (uncooled detectors) and SNSPDs (cooled detectors).
\section{Methods}
We consider 4 solutions to the problem of generating a fully connected untrusted node network: uncooled Decoy BB84, cooled Decoy BB84, uncooled TF-QKD and cooled TF-QKD. Only the cooled TF-QKD solution has detectors localised on a few nodes. \par
We consider a network defined by $G=(V,E)$ where the vertices are split in two sets, nodes that wish to generate secure keys with each other $S \in V$ and nodes that have been specified as potential locations for detectors $C \in V$. This mirrors a realistic scenario where a company has some locations that can cheaply be converted to buildings capable of storing multiple detectors. $N_{bob}$ of the nodes in $C$ are turned `on': $B \in C$, specifying how many locations the company is willing to convert into detector hubs.\par
To optimise the locations for every permutation of `on' detectors, the algorithm calculates the minimum distance of every source node to every `on' detector, which will result in the optimum rate for that detector as shown in \textbf{\cite{Wang2020}}. A list of these distances $(i,b,L_{i,b}^{min})$ is kept. For every pair $i,j \in S, i<j$, we use the lengths and formulas described in section 2 to calculate the secure key rate for each `on' detector node $c_{i,j}^{max, b}$. The capacity of the link is then $c_{i,j}^{max} = \max_{ b \in B} c_{i,j}^{max,b}$. We look at optimising the total capacity
\begin{equation}
    c_{net}^{B} = \sum_{i,j \in S: i < j} c_{i,j}^{max}
\end{equation} \par
For the Decoy BB84 protocol, no localisation is possible. In order that all nodes in $S$ can communicate with each other in an untrusted node topology, each node must contain a source and a detector. Here we assume there is a source and a detector for each connection, which is the same assumption made in the TF-QKD optimisation. We consider the same graph $G = (V,E)$ and find the shortest path distance $L_{i,j}^{min} ~ : \forall i,j \in S$. From this, we calculate the capacity of the connection $c_{i,j}$ using SPADs and SNSPDs and then the network capacity is $c_{net}^{Decoy} = \sum_{i,j \in S: i \neq j} c_{i,j}^{max}$. Similarly, for the uncooled TF-QKD solution, we assume it is inexpensive to place the detectors in the midpoints of the path. Thus, the same method is used to calculate the rates here, with the formula for the TF-QKD rates used instead.\par
\begin{figure}[t]
   \centering
    \includegraphics[width=0.6\linewidth]{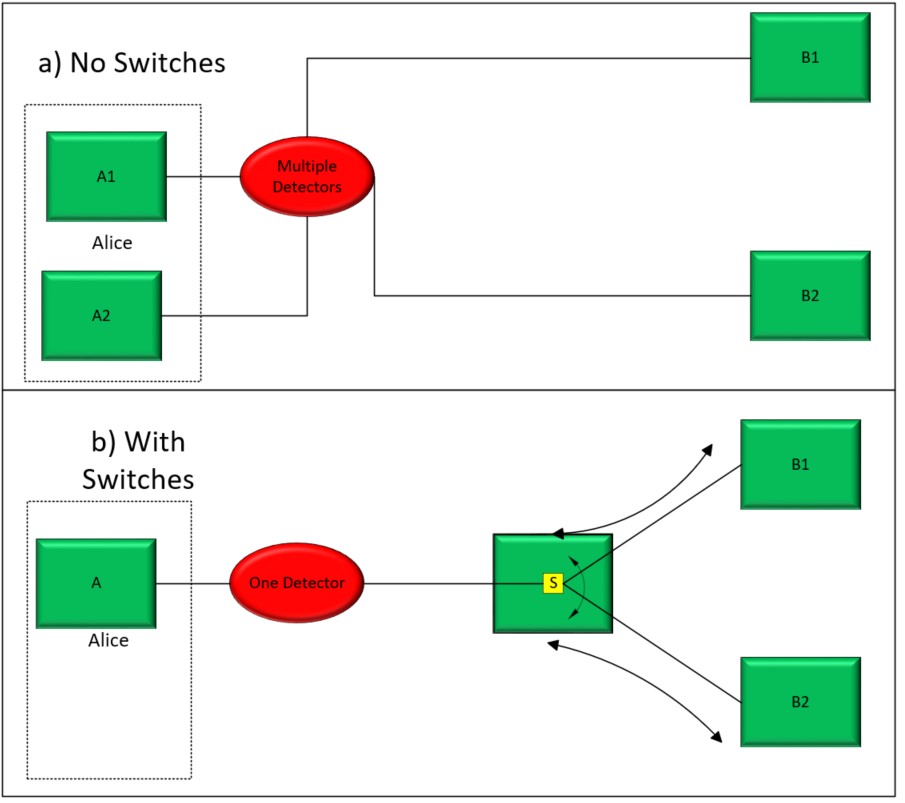}
    \caption{
    a) The setup for a simple network without switches. There are a lot of fibres needed to ensure connectivity and also a large number of detectors and sources to allow Alice to communicate with both B1 and B2. b) The same network using switches. Far fewer fibres are used and only one source and detector are necessary for Alice to communicate with both B1 and B2.}
    \label{fig:figureswitches}
\end{figure}

In the second half of the investigation, we investigate the effect of adding the losses introduced by the switches to the model. In a realistic implementation, switches will be necessary to route the signals. Figure \ref{fig:figureswitches} shows the benefits of switches in a simple network. Without switches, one fibre resource, i.e. one wavelength on a fibre, needs to be reserved for each connection and one detector and source are needed for each connection. With switches, only one detector/source pair and one fibre resource is needed for the network in figure \ref{fig:figureswitches}. This reduces network costs drastically. The switches have an associated 1-2dB loss, which must be taken into account when considering the number of nodes that will need to be traversed due to localisation. We add this loss into the graph and carry out the same analysis as in the first half of the investigation.


\section{Results}
\subsection{No Switches}
\begin{table*}
  \centering
    \begin{tabular}{|c|P{1.5cm}|c|c|c|}
        \hline Solution & 0 Capacity Size/Km & \multicolumn{3}{|c|}{Total Capacity Ratio: TF-QKD Cooled / Current Solution}  \\
        \hline
         No. of Graph Nodes&  $|S| = 40$  & $|S| = 20$ & $|S| = 30$ &  $|S| = 40$ \\
        \hline
        Decoy BB84 Uncooled & $50$ & $34 \pm 3$ & $33 \pm 3$ & $32 \pm 3$ \\
        \hline
        Decoy BB84 Cooled & $80$ & $0.92 \pm 0.08$ & $0.90 \pm 0.08$ & $0.87 \pm 0.07$ \\
        \hline
        TF-QKD Uncooled & $120$ & $3.7 \pm 0.6$ & $3.4 \pm 0.5$ & $3.1 \pm 0.3$\\
        \hline
        TF-QKD Cooled & $110$ &  - & -&  -\\
        \hline
    \end{tabular}
    \caption{Size of the network for which a solution can be found that is fully connected with no connection having 0 capacity and the ratio of capacities to TF-QKD Cooled Solution proposed for graphs with $|C| = 10, |B| = 4$ and average connectivity per node 3.5 in a 100km box.}
    \label{tab:results}
\end{table*}
\begin{figure}[t]
  \centering
  \begin{subfigure}[t]{0.45\textwidth}
         \centering
         \includegraphics[width=\textwidth]{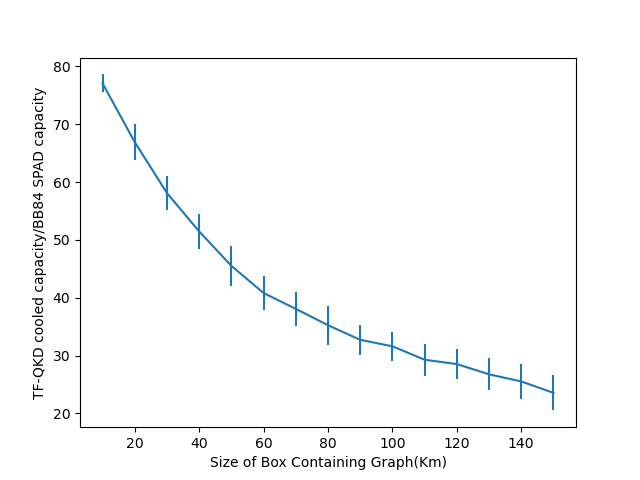}
         \label{fig:}
     \end{subfigure}
     \begin{subfigure}[t]{0.45\textwidth}
         \centering
         \includegraphics[width=\textwidth]{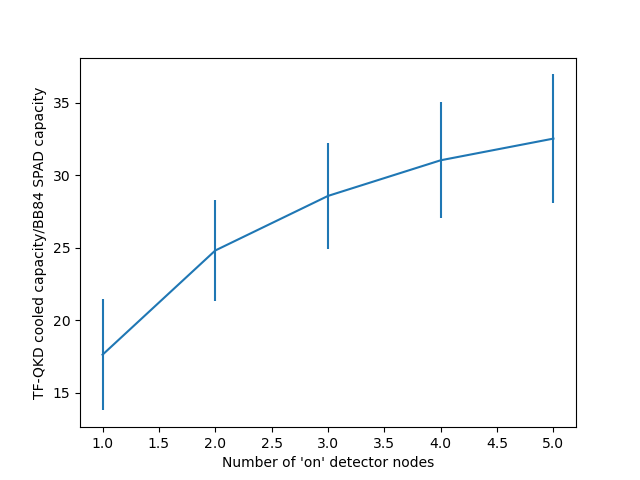}
         \label{fig:}
     \end{subfigure}
     \caption{Trends of fraction of rates of TF-QKD cooled solution capacities vs BB84 hot capacity for different network sizes (left) and different number of on detector nodes in the network (right)}
    \label{fig:figure5}
\end{figure}
Table \ref{tab:results} presents the results for the different solutions for different number of nodes in the graph in the absence of switch losses. To obtain the results, we generate 50 random graphs, find the optimum solution for that graph and then average the results over all the graphs. For the size of the network for which the solution can generate a fully connected solution we use the largest size for which $< 10\%$ of the graphs have at least one connection with $0$ capacity, by reasoning that a small number of $0$ capacity connections are possible for small graphs with some topologies. This was justified experimentally, with rare cases of graphs yielding a 0 capacity edge at box sizes much smaller than the majority of graphs having 0 capacity. In real networks, some tailoring to optimal locations of detectors will be made while in the current investigation they are placed randomly and thus it is possible to get all nodes on the edges of the box. The method of selection is thus reasonable under the assumption that reasonable locations are selected as possible detector nodes. A uniform method of determining the maximum size of fully connected solutions ensures no bias in the results, considering the capacities are calculated for all solutions on the same graphs.\par
From Table \ref{tab:results}, it is clear that TF-QKD drastically improves the overall size of possible fully connected networks as compared to Decoy BB84. This is a result of the extended range of the protocol. Interestingly, despite localisation of the detectors, the cooled TF-QKD solution can generate fully connected network solutions at a distance only marginally lower than the uncooled TF-QKD protocol. This shows that cooling the detectors is sufficient to overcome the loss of range from localisation of the detectors. This is remarkable, since the uncooled TF-QKD solution is unfeasible when adding more nodes to the graph, as this requires changing the entire infrastructure of the network, but similar rates and distances are possible by localising and cooling the detectors.\par
We also see that the average capacity per connection, characterised by the total capacity ratio, is significantly improved from the uncooled Decoy BB84 protocol and is similar to the cooled Decoy BB84 and to the uncooled TF-QKD solutions. The latter is likely a result of the high capacity, short distance connections, as can be seen from figure \ref{fig:figure2}, with the capacity of the cooled TF-QKD being $\sim 10 \times$ larger than those for the uncooled TF-QKD at $>100 km$. However, this is not true for the cooled Decoy BB84 protocol and the similarity in average rate reflects the fact that the connection rate is not significantly altered by localisation. There is a slight reduction in the ratio with increasing number of nodes, indicating that in the case of bigger graphs the mean capacity of the TF-QKD is reduced slightly compared to those where localisation is not used. This reflects the increased distance of connections associated with the increased number of node hops needed per connection due to detector localisation. However, this reduction is slight with variations over graphs being more significant.\par
Figure \ref{fig:figure5} illustrates the trends found in the simulations. In particular, we see that the benefits of TF-QKD cooled solution decreases as compared to other solutions in longer distance networks, however there is still significant benefit even at large distances. This relative decrease in total rate of the cooled TF-QKD compared to other solutions is likely attributed to the fact that localisation in larger networks has a larger effect on the capacities, due to the exponential decrease in rates with distance. We also see that adding more detector locations increases the overall performance of the network. However, in this case, the increase in cost of the network by adding cooling locations may mean this increase in performance is not worth the added cooled nodes. It was found that changing the average number of connections per node in the graph did not change the ratios of the total capacities.

\subsection{Switches}
Table \ref{tab:resultsswitches} presents the results for the different solutions in the presence of switch losses for typical values of such losses. The ratios of the overall rates of TF-QKD cooled solution to the overall rates of other solutions are reduced compared to those calculated without switch losses, showing that the switch losses affect the localised detector solution more than direct connections. This was expected, as localisation of the detectors increases the average number of node hops needed for a connection, and thus the overall loss due to switches. In contrast, the switch losses does not reduce the size of the network which is possible for the cooled TF-QKD more significantly than the non-localised solutions, with the size being only marginally smaller than the uncooled solution. The possible size of the network looks at the highest loss connections, which often contain a significant number of node hops. While localisation does increase the number of node hops needed, the fractional increase is small. Further, as high-capacity connections have reduced capacities due to the loss from switches, the placement of detectors will be more appropriate for long distance connections. The combination of these two effects is that the size of the network possible follows similar trends to the uncooled, unlocalised solution. We also see that there is a significant advantage in scale from using the TF-QKD localised detector solution compared to a BB84 solution. It is important to note that these values are calculated for graphs with $|S| = 40$, and with more nodes the area of fully connected solutions is further decreased. It can be seen that the Decoy BB84 uncooled solution cannot generate keys on graphs with this many nodes for typical switch losses, while even a fully cooled solution is only capable of generating a fully connected network solution for $1dB$ losses at short distances. The significant drop in range when increasing the losses in switches reinforces the need for manufacturers to stick to $1dB$ or less loss switches for quantum technologies.

\begin{table*}
  \centering
    \begin{tabular}{|c|P{0.8cm}|P{0.8cm}|P{0.8cm}|P{2cm}|P{2cm}|P{2cm}|}
        \hline Solution & \multicolumn{3}{|c|}{0 Capacity Size/Km} & \multicolumn{3}{|c|}{Total Capacity Ratio: TF-QKD Cooled / Current Solution}  \\
        \hline
         Switch Loss (dB) &  1 & 1.5 & 2  & 1 & 1.5 &  2 \\
        \hline
        Decoy BB84 Uncooled &$<10$ & $<10$ & $< 10$ & $28 \pm 4$ & $26 \pm 4$ & $27 \pm 4$ \\
        \hline
        Decoy BB84 Cooled & 20 & $<10$ & $<10$ & $0.76 \pm 0.13 $& $0.70 \pm 0.12$& $0.73 \pm 0.12$\\
        \hline
        TF-QKD Uncooled & 80 & 70 & 40 & $2.3 \pm 0.4$ & $2.1 \pm 0.4$ & $2.1 \pm 0.4$\\
        \hline
        TF-QKD Cooled  & 70 & 60 & 30 & -& -& -\\

        \hline
    \end{tabular}
    \caption{Size of the network for which a solution can be found that is fully connected with no connection having 0 capacity and the ratio of capacities to TF-QKD Cooled Solution proposed for graphs with $|S| = 40, |C| = 10, |B| = 4$ and average connectivity per node 3.5 in the presence of switches at each node with different losses in the typical range. Ratios given for 100km boxes for fair comparison.}
    \label{tab:resultsswitches}
\end{table*}
\section{Conclusion}
In this paper, we propose taking advantage of the TF-QKD topology, which allows for the collocation of detectors in a network to take advantage of the very high detection efficiencies of SNSPDs whileat the same time keeping the costs of cooling these detectors low. This allows for an affordable, substantial improvement of the achievable rates and distances for which it is possible to generate quantum keys in a network. We use computer simulations to show that the proposed method greatly increases the scale of possible fully connected networks compared to Decoy BB84. With just 4 cooling locations, TF-QKD is capable of a fully connected solution up to $110km$ sided square for a $40$ node network, whereas Decoy state BB84 with all nodes cooled is only able to extend to $80km$. Furthermore, we find that the decrease in range of the network due to localisation is offset by cooling the detectors, with the uncooled TF-QKD solutions, which does not have localised detectors, but rather the detectors are placed at the midpoint of the connections, only able to generate fully connected solutions up to $120km$, a small increase in range given the significantly increased cost for such a solution. We also found that the average rates of the localised, cooled TF-QKD solution are $\sim 0.9$ times those rates generated by a cooled Decoy BB84 solution and $>30$ times improvement to the uncooled Decoy BB84 solution.\par
When adding switch losses into the model, which reduces the cost of the network further, we see a drastic improvement in range of the network for TF-QKD compared to the Decoy BB84, with only cooled Decoy BB84 being capable of generating a fully connected solution on a $40$ node network and even then only up to $20km$ compared to the $70km$ range of the localised cooled TF-QKD with $1dB$ switch losses. Thus, the conclusion that there is only a minor decrease in range from localisation, when cooling is added, still holds in the presence of switches with the range of localised, cooled TF-QKD being $10km$ shorter than that for the unlocalised, uncooled TF-QKD independent of the switch loss. This study shows that TF-QKD is appropriate for network scenarios, allowing for a low cost, substantial improvement to the previous Decoy BB84 solution. 

\section*{Acknowledgements}
This work was supported by the UK Engineering and Physical Sciences Research Council (EPSRC) EP/V519662/1 and by BT. The modelling was done using graph-tool \cite{peixoto_graph-tool_2014}.
\bibliography{references}
\section*{Appendix A}
In this appendix, we will describe the method used to calculate the values of the secure key rates given by equation (2). As explained in the main text we use methods from \cite{Wang2020}, however it is necessary to clarify certain assumptions.\par
We will use the same notation as in the paper \cite{Wang2020}. In particular, for channel intensities $s_A$ and $s_B$ with transmittance between Alice and Bob to Charlie $\eta_A, \eta_B$, let $$\gamma _i = s_i \eta_i ~~ \forall i \in \{A,B\}$$ and for the decoy states, with intensity for the $j^{th}$ decoy state $\mu_{k}^{j}$: $$\gamma_{k}'  = \mu^{j}_k \eta_k ~~\forall k \in \{A,B\}$$ Here we use the value $s_i = 0.1^2$ for the strongest of the two signals.
Using these definitions \cite{Wang2020} defined $$p'_{XX}(k_c,k_d)= \frac{1}{2}(1-p_{DC})(e^{-\sqrt{\gamma_A\gamma_B}\cos \phi\cos\theta} + e^{\sqrt{\gamma_A\gamma_B}\cos\phi\cos\theta}) e^{-\frac{1}{2}(\gamma_A + \gamma_B)} - (1-p_{DC})^2e^{-\gamma_A-\gamma_B}$$ where the prime is used to indicate the detector dead time has not been accounted, and $\theta$ is the polarisation misalignment between Alice and Bob and $\phi$ is the phase misalignment. Here we assume the decoy states are sent rarely and thus do not affect the signals significantly in regards to the dead time. As $p_{XX}(1,0)$ and $p_{XX}(0,1)$ represent clicks on different detectors used in the interference measurements the different click results will not affect each other and thus we can include the dead time with $$p_{XX}(k_c,k_d) = p'_{XX}(k_c,k_d)\eta_{dead}(k_c,k_d)$$ $$\eta_{dead}(1,0) = (1+\tau_{dead}f_{rep}(p'_{XX}(1,0)+p'_{XX}(1,1)))^{-1}$$
$$\eta_{dead}(0,1) = (1+\tau_{dead}f_{rep}(p'_{XX}(0,1)+p'_{XX}(1,1)))^{-1}$$
From the Hong-Ou-Mandel effect \cite{Hong1987}, we know that the probability that both detectors click in the scenario where we have at most one photon into each beam is 0 as the photons bunch, as long as the photons are indistinguishable. This means double clicks come from terms that involve multiphoton states or decoherence. The first condition is an order of magnitude less than the probability that the detectors click and can thus be ignored due to the low intensity of the field used. The decoherence becomes important in the long distance limit where there is significant decoherence of the state. The probability of a detection event occurring during the dead time is small in this limit and thus this term is not the limiting factor. Alternatively, the decoherence term can be though of as resulting from the loss in visibility of the photons. Using the same formula as the Decoy BB84 $$ V = \frac{\mu \eta_f\eta_{d}}{\mu \eta_f \eta_{d} + 2P_{e}}$$ we see in the short distance limit $\eta_f \approx 1, \eta_d \approx 0.2-0.8, \mu = 0.1$, thus $\mu\eta_f\eta_d >> P_e$ and the visibility of the photons is $V \approx 1$ so can be ignored in the short distance limit. In fact, from this formula we can determine an approximate distance above which the decoherence becomes important. In particular, $\frac{P_e}{\eta_d\mu} \approx 10^{-5}$. So we consider there to be a significant effect when $\eta_f \approx 10^{-4}$. This occurs at around $200km$ above which the dead time is not a limiting factor.  Thus the $p'_{XX}(1,1)$ can be ignored in the dead time calculation without significantly affecting the results. Therefore, we use $$p_{XX}(k_c,k_d) = p'_{XX}(k_c,k_d)\eta_{dead}(k_c,k_d)$$ $$\eta_{dead}(1,0) = (1+\tau_{dead}f_{rep}p'_{XX}(1,0))^{-1}$$
$$\eta_{dead}(0,1) = (1+\tau_{dead}f_{rep}p'_{XX}(0,1))^{-1}$$ 
The QBER is given by $$e_{X,k_ck_d} = \frac{e^{-\sqrt{\gamma_A\gamma_B}\cos\phi\cos\theta} -(1-p_{DC})e^{-\frac{1}{2}(\gamma_A + \gamma_B)}}{e^{-\sqrt{\gamma_A\gamma_B}\cos\phi\cos\theta} + e^{\sqrt{\gamma_A\gamma_B}\cos\phi\cos\theta} - 2(1-p_{DC})e^{-\frac{1}{2}(\gamma_A + \gamma_B)}}$$

The upper bound phase error is given by \cite{Wang2020} $$e^{upp}_{Z,k_ck_d} = \sum_{j=0,1} [\sum_{m_A,m_B = 0}^{\infty} c_{2m_A+j}^{A, (j)}c_{2m_B+1}^{B,(j)}\sqrt{p_{ZZ}(k_c,k_d|2m_A+j, 2m_B+j)}]^2$$ where $c_
{2n}^{A,(0)} = \frac{\alpha^{2n}_A}{\sqrt{2n!}}$, $c_{2n+1}^{A, (1)} = \frac{\alpha^{2n+1}_A}{\sqrt{2n+1!}}$, $c_
{2n}^{B,(0)} = \frac{\alpha^{2n}_B}{\sqrt{2n!}}$, $c_{2n+1}^{B, (1)} = \frac{\alpha^{2n+1}_B}{\sqrt{2n+1!}}$ and all other combinations are 0. We assume the infinite decoy state, due to calculation times, where \cite{Wang2020} $$p'_{ZZ}(k_c,k_d|n_A,n_B) = (1-p_{DC})q_{ZZ}(k_c,k_d|n_A,n_B) +(1-p_{DC})p_{DC}(1-\eta_A)^{n_A}(1-\eta_B)^{n_B}$$
\begin{align*}
    q_{ZZ}(k_c, k_d|n_A, n_B) & = \sum_{k=0}^{n_A}\binom{n_A}{k}\sum_{l=0}^{n_B}\binom{n_B}{l} \frac{\eta_A^{k}\eta_B^l(1-\eta_{A})^{n_A-k}(1-\eta_B)^{n_B-l}}{2^{k+1}k!l!}\sum_{m=0}^{k}\binom{k}{m}\sum_{p=0}^{l}\binom{l}{p} \\
     & \sum_{q=\max(0,m+p-l)}^{\min(k,m+p)}\binom{k}{q}\binom{l}{m+p-q} (m+p)!(k+l-m-p)!\cos^{m+q}(\theta_A) \\
      &\cos^{m+p-q}(\theta_B)\sin^{2k-m-q}(\theta_A)\sin^{2l-m-2p+q}(\theta_{B})-(1-\eta_A)^{n_A}(1-\eta_B)^{n_B}
\end{align*}

where $\theta_i: ~ i \in \{A,B\}$ is the phase misalignment of the senders with respect to Charlie, and $$p_{ZZ}(k_c,k_d|n_A,n_B) = p'_{ZZ}(k_c,k_d|n_A,n_B)\eta_{dead}(k_c,k_d)$$ We note that only the terms up to $n_A=n_B = 5$ are calculated for the calculation of $e^{upp}_{Z,k_ck_d}$ and the rest of the terms $p_{zz}(k_c,k_d|2n_A+j,2n_B+j) = 1 ~\forall n_A,n_B > 5$.

\end{document}